\documentclass[lettersize,journal]{IEEEtran}
\usepackage{amsmath,amsfonts}
\usepackage{amssymb}
\usepackage{algorithmic}
\usepackage{algorithm}
\usepackage{array}
\usepackage{amsthm}
\usepackage[caption=false,font=normalsize,labelfont=sf,textfont=sf]{subfig}
\usepackage{textcomp}
\usepackage{stfloats}
\usepackage{url}
\usepackage{verbatim}
\usepackage{graphicx}
\usepackage{cite}
\usepackage{xcolor}
\usepackage{multirow}
\usepackage{fancyhdr} 
\usepackage[justification=centering]{caption}
\def\BibTeX{{\rm B\kern-.05em{\sc i\kern-.025em b}\kern-.08em
    T\kern-.1667em\lower.7ex\hbox{E}\kern-.125emX}}

\begin{document}

\title{REMR: A Reliability Evaluation Method for Dynamic Edge Computing Network under Time Constraints}

\author{Liang Chen, Jianpeng Qi, Xiao Su, Rui Wang \\
    \IEEEmembership{This work has been submitted to the IEEE for possible publication. Copyright may be
        transferred without notice, after which this version may no longer be accessible.}
    \thanks{This paper was produced}
    \thanks{Manuscript received April 19, 2021; revised August 16, 2021.}}

\markboth{Journal of \LaTeX\ Class Files,~Vol.~14, No.~8, August~2021}%
{Liang \MakeLowercase{\textit{et al.}}: A Sample Article Using IEEEtran.cls for IEEE Journals}


\maketitle

\begin{abstract}
    While the concept of Artificial Intelligent Internet of Things\ (AIoT) is booming, computation  and/or communication-intensive tasks accompanied by several sub-tasks are slowly moving from centralized deployment to edge-side deployment. The idea of edge computing also makes intelligent services sink locally. But in actual scenarios like dynamic edge computing networks (DECN), due to fluctuations in available computing resources of intermediate servers and changes in bandwidth during data transmission, service reliability becomes difficult to guarantee. Coupled with changes in the amount of data in a service, the above three problems all make the existing reliability evaluation methods no longer accurate. To study the effect of distributed service deployment strategies under such a background, this paper proposes a reliability evaluation method (REMR) based on lower boundary rule under time constraint to study the degree of the rationality of a service deployment plan combined with DECN. In this scenario, time delay is the main concern which would be affected by three quantitative factors: data packet storing and sending time, data transmission time and the calculation time of executing sub-tasks on the node devices, specially while the last two are in dynamic scenarios. In actual calculation, based on the idea of the minimal paths, the solution set would to be found  that can meet requirements in the current deployment. Then the reliability of the service supported by the solution sets would be found out based on the principle of inclusion-exclusion combined with the distribution of available data transmission bandwidth and the distribution of node available computing resources. Besides a illustrative example was provided, to verify the calculated reliability of the designed service deployment plan, the NS3 is utilized along with Google cluster data set for simulation. The results gained from NS3 proved the accuracy and applicability of the proposed reliability evaluation method.
\end{abstract}


\section{Introduction}
\IEEEPARstart{W}{hen} a large number of devices compositing the Internet of Things entered  people's lives in various scenarios including smart cities, smart industries and smart homes, the era of the Internet of Everything comes. Gartner said that 75\% of enterprise data would be generated on the edge by 2022 \cite{top10strategic}, and the IDC believes that 41.6 billion edge devices would be interconnected by 2025, with the data volume close to 80ZB \cite{shirer2019growth}. While the new concepts of Industrial Internet of Things, Urban Internet of Things, Home Internet of Things and so on, bring opportunities \cite{nord2019internet}, there are also countless challenges. When computation and/or communication-intensive tasks  get gradually closer to the edge \cite{8736011}, one of the challenges is that the data processing capability of the Internet of Things as a perception network has never been comparable to the data center. Then cloud computing just makes up for this shortcoming. This computing paradigm combines the Internet of Things and cloud computing with the Internet. The devices on the edge would collect data and send them to the cloud for computation and storage. But the cloud computing is not that perfect.

\begin{figure}[htbp]
    \centerline{\includegraphics[width=0.5\textwidth]{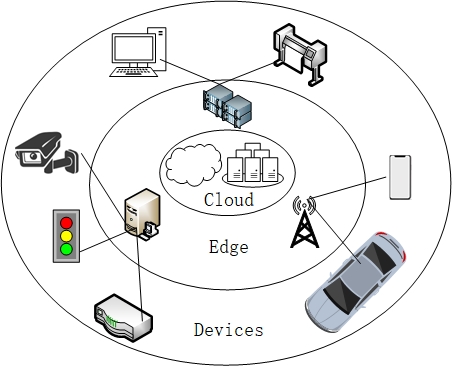}}
    \caption{Typical edge computing examples}
    \label{fig1.1}
\end{figure}

Although centralized cloud services like Amazon Web Service (AWS) helped companies effectively reduce IT investment costs and maintenance costs \cite{robinson2014using}, their shortcomings are also obvious. Limited network transmission performance, high energy consumption and maintenance costs, difficulty for privacy guarantee, and poor real-time quality, have yet to be resolved. Edge computing emerged in this situation. Edge topics have appear in related research fields since 2015 \cite{shi2016edge}, such as Mobile Edge Computing (MEC) \cite{8030322}, micro-cloud and fog computing \cite{yi2015survey}. Compared with cloud computing owning excellent computing capabilities, edge computing uses nearby services to provide faster service response, simultaneously realizes effective use of local resources showed in Fig.\ref{fig1.1}. Edge devices are generally considered to own insufficient computing capabilities \cite{king2016distributed}, which is in line with reality that edge devices would not execute difficult computation task. The general way works that setting edge servers between cloud computing and edge devices to complete the computing tasks of the data transmitted from the devices instead of entirely relying on the cloud computing center \cite{mao2017survey}. Through this mode, only part of the intermediate data or the final calculation results need to be transmitted to the cloud, which greatly reduces the cost of communicating with the cloud. It is unreasonable for the camera at the gate of a community to recognize someone passing by without auxiliary equipment. This task requires an additional VGG-Net \cite{simonyan2014very} or any model designed for face recognition. Obviously, the computing capability does not support it to complete this task. But it is possible to perform preliminary processing on the captured pictures and videos. A complete VGG16 model with 138M parameters will take up close to 500Mb of disk space, but the pre-process operation will not require so many parameters. A service completed in the form of multiple sub-services, could improve the application’s scalability, portability, and availability.\cite{fu2021adaptive}

Edge computing, this computing paradigm has present outstanding performance and practicality, while many researchers have published relevant work \cite{satyanarayanan2017emergence} \cite{9272869 }, there exists still two issues. Without cloud computing, the network transmission latency brought by high-volume data transmission is reduced, but the realization of most computing work outside the cloud center would cause the computational latency. The edge servers and devices cannot rival with cloud on computing resources, so the computational delay becomes an issue that must be considered when designing and deploying distributed services. The next is that the stability of cloud computing mode disappeared. Due to redundant configuration, cloud computing guarantees that the requests from the devices can always be completed attributing to sufficient computing resources. But when combining distributed services with edge computing scenarios, it is necessary to consider the issue of computing resources. Specially each edge server is generally responsible for more than one service, which means that its available resources are changeable with time \cite{milocco2020evaluating}. When a new task arrives, the amount of  currently available computing resources of the server is unknown, making it difficult to estimate the execution time. And the available network bandwidth is always changing, which also make the entire time cost fluctuating. The execution time of the task becomes difficult to estimate due to these uncontrollable factors, and the originally designed service is likely to become an unreliable service, making it difficult to realize a major advantage of edge computing. For example, short-term network congestion or insufficient machine performance will lead to a decrease in quality of service (QoS).

In addition, the change in the amount of data in transmission process of distributed tasks, which means the amount of data that needs to be transmitted at one service is different, also makes the existing reliability evaluation methods inapplicable \cite{huang2020network}. To solve the dilemma, this paper models the distributed tasks and deployment plan in the above scenario based on the concept of reliability, built a DECN scenario and  proposed a method to calculate the probability that a distributed task can be completed, also named reliability. Aiming at the unreliability of bandwidth, bandwidth distribution sampling was used to represent the change in available bandwidth. In view of the changes in computing resources of computing nodes, the probability distribution of available computing resource is summarized based on historical data modeling. Combining the characteristics of distributed tasks, find the solution set that meets the requirements to calculate the reliability of completion. The proposed method can not only tell the credibility of the current deployment plan to complete the assigned tasks, but also give a theoretical support in the early stage to help service providers deploy a more perfect service.

The detailed contributions of this paper are as follows:
\begin{itemize}
    \item This study provides a brand new reliability evaluation method aiming at popular edge computing scenarios with complicated dynamic edge computing networks. To our best known, it is the first try that network bandwidth and computing resources distribution are jointly considered to evaluate the reliability of edge computing distributed services under the complex dynamic scenarios with changeable data size.
    \item Different from traditional network reliability evaluation, dynamic changes in the amount of data on the deployment plan and the computing  performance of nodes are considered and implemented.
    \item This study achieved static separation of composite node and transmission node considering the actual scenario, which means that the proposed DECN can fit nodes with different functions.
\end{itemize}
In the rest of this paper, section 2 introduced related works about reliability in network transmission and edge computing. Theoretical support and basic experimental procedures were presented in section 3, along with a reliability algorithm. An illustrative example and simulation were given in Section 4 to bring more convincing details. Section 5 concluded the whole study and gave some directions in future research, focusing on the content that has not been explored in depth in this paper.

\section{Related works}
\begin{figure}[htbp]
    \centerline{\includegraphics[width=0.45\textwidth]{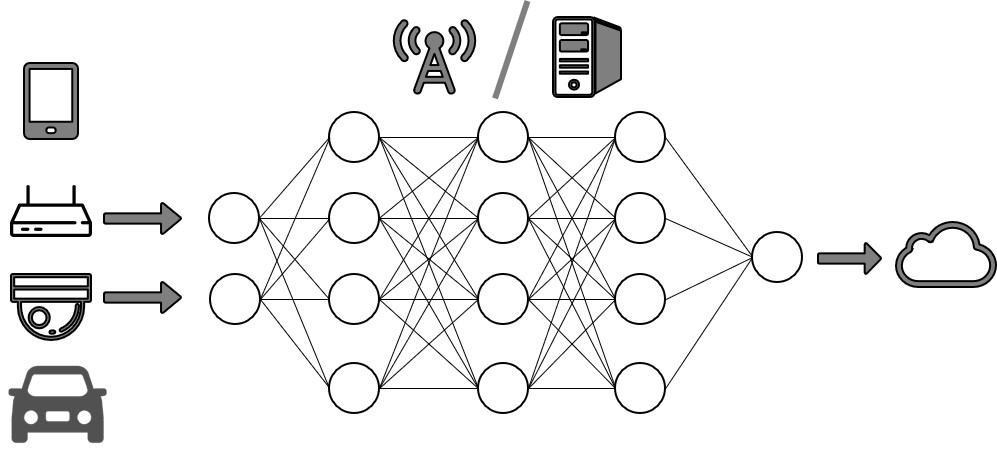}}
    \caption{Edge computing combined with deep learning methods}
    \label{fig2.1}
\end{figure}

The traditional reliability often appeared in related manufacturing and supply chain researches and applications \cite{8576668}. While the focus on reliability is shifted to computer-related research, it is found that many researchers in network transmission use it \cite{8776630}, called ``networks reliability analysis''. When combining edge computing with deep learning services like Fig. \ref{fig2.1}, the reliability index of the system also needs to be considered. Then the reliability discussed in this article is different also, which is used to express success probability of distributed services in edge computing.

\subsection{Networks Reliability}
Researchers have done a lot of work on the reliability of network transmission \cite{huang2020network} \cite{8932578}. Paper \cite{7964780} uses the concept of elasticity, which combines network capacity and reliability that affect service quality. The proposed calculation method more comprehensively considers the influencing factors of service reliability, uses rerouting to solve failures and splitting data to solve network congestion. When Lin is modeling the cloud and fog system, the proposed stochastic-flow cloud–fog computing network model effectively evaluates the probability of successful data transmission in the IoT scenario \cite{9468876}. Since their model ignored the computing process of nodes and the dynamic computing resources of nodes, their method cannot be directly applied to edge computing scenarios with distributed tasks. Ruen et al. used an approximate Lagrangian relaxation algorithm to solve the problem of generating the maximum reliability of a computer network under a delay constraint \cite{loh2011addressing}. In paper \cite{huang2021novel}, Lin et al. creatively applied deep neural network to the calculation of network reliability and gained promising performance.

However in addition to thinking of the minimum boundary required for data transmission, it is also necessary to consider the minimum boundary required for task when computing resources are dynamically changing. Combined with minimal cut vectors, Ramirez-Marquez proposed the Monte Carlo simulation method to estimate the reliability of multi-state networks, and achieved better numerical results than enumeration method, inclusion and exclusion principles \cite{ramirez2005monte}. It provides an effective way to solve the problem from other directions. However, in certain aspects this simulation method still lacks sufficient applicability, taking into account the calculations that may exist in the network. It is similar from the perspective of providing services and completing tasks. But in the scenario of edge computing, in addition to data transmission, computing is also an important point which need to be considered. So measuring the reliability of a service requires comprehensive consideration, especially when computing resources and network bandwidth fluctuate.

\subsection{Edge Computing Reliability}
When Dong et al. designed the optimal offloading strategy and allocation algorithm in the resource-constrained mobile edge computing system \cite{dong2019reliability}, they considered the reliability of the task which was based on the inert shadow scheme. The way that relying on redundancy deployment to improve reliability is commonly used, like the disaster tolerance performance of the server services. Li et al. are studying task allocation of the edge computing system composed of ES and EU \cite{li2019edge}. Their model selectively offload part of the tasks to the local devices to complete the calculation, and the remaining part will be calculated by ES with regular model retraining and parameter return. However, it is assumed that the local model will execute at a fixed CPU frequency, and the dynamic change of the available resources of the devices and edge server are not taken into account. The delay and energy consumption results obtained in this way are lacking of some important consideration. Kang and others have questioned the current pure cloud computing paradigm in paper \cite{kang2017neurosurgeon}. By comparing the pure cloud and cloud-edge collaborative computing paradigm, they explained the importance of sinking part of the computing to the edge. And proposed a framework, which can adaptively cut the model at a reasonable division point and send them to different computing units with the goal of lowest delay and  energy consumption. There is no doubt that the work of their thesis is very representative. But the work does not take the dynamic factors of resources into consideration, which means that it achieves a limited lowest delay only in the reasonable division of the analysis task in a static way. Compared with the above work, in a typical edge computing scenario, REMR not only combines the network transmission delay and the calculation delay of different nodes, but also realizes dynamic modeling according to the fluctuation of network bandwidth and the availability of nodes resources. on this basis, the proposed reliability evaluation method can effectively offer the feasibility of the current deployment plan, and it has been verified on the simulation platform NS3 that the calculation results are almost the same as the simulation results.

\section{Methods and Materials}
\begin{figure}[htbp]
    \centerline{\includegraphics[width=0.5\textwidth]{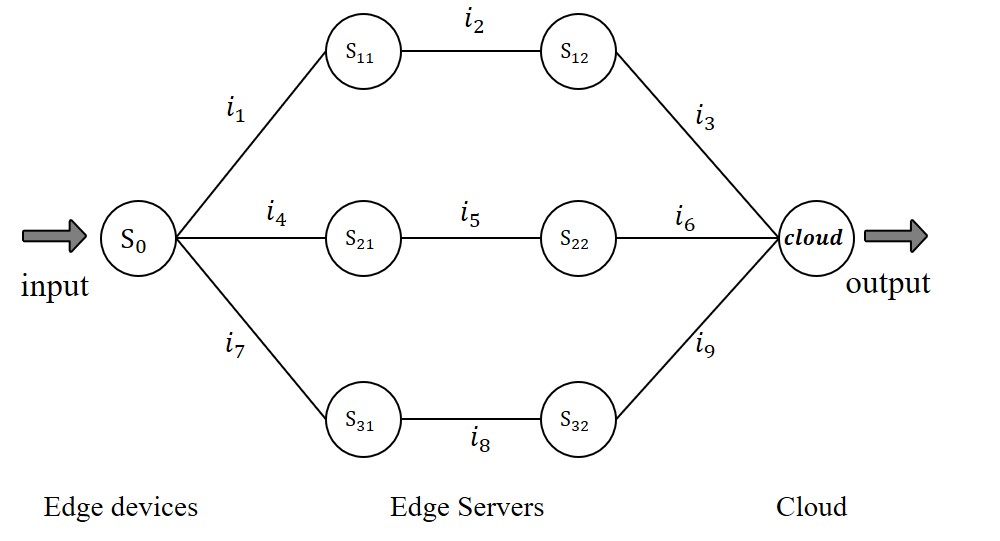}}
    \caption{An brief DECN scenario}
    \label{fig3.1}
\end{figure}

In DECN, a distributed service or application is deployed on the device side and edge servers that can provide computing resources according to the corresponding strategy like Fig. \ref{fig3.1}. To improve the reliability of global task completion, there usually exists several deployment plans in one DECN situation. After the service deployment finished according to the hierarchical structure, the model assigned to the devices will process the input data, and the intermediate data obtained will be transmitted to the edge server through the line for the next step of the whole service. For calculations, depending on the distribution of models and the amount of calculations, different deployments may have different numbers of intermediate servers. In the end, the results of all deployments will be transmitted to the cloud, which mainly plays a role of monitoring and recording. There may also exist some servers for transmission only.

\begin{figure}[htbp]
    \centerline{\includegraphics[width=0.5\textwidth]{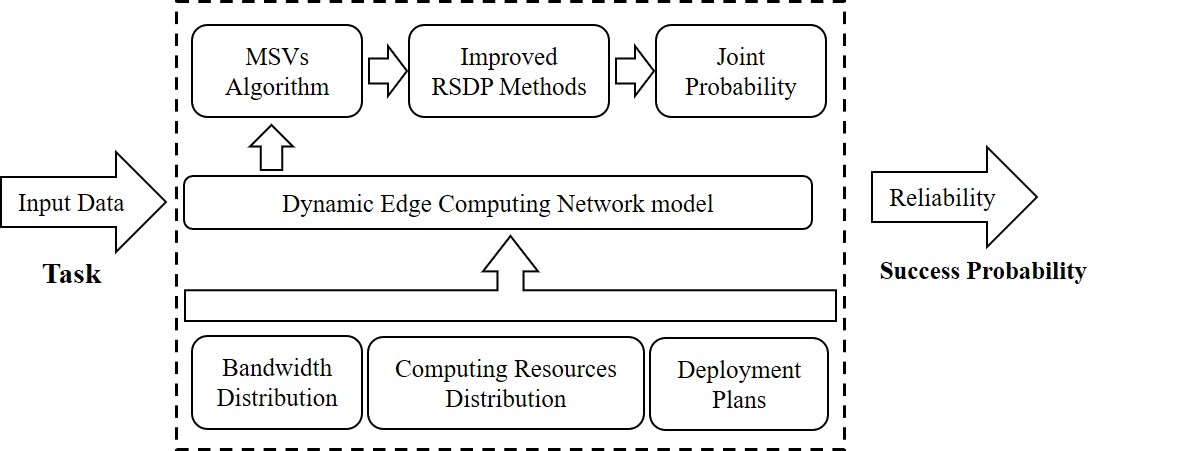}}
    \caption{The workflow of the DECN model}
    \label{fig3.2}
\end{figure}

Similar to traditional network transmission, each transmission line in one deployment  has a possible capacity and corresponding probability distribution. The difference is that the traditional network transmission model is no longer applicable here because the middle servers need to complete a large number of computing tasks. Therefore, for intermediate service providers, additional considerations have been taken. combining the characteristics of edge servers and considering the services that may have be running, the computational performance is modeled in a dynamic mode, and expressed in the form of probability distribution. Due to the randomness of transmission capacity and servers computing performance, the successful execution of the current task becomes a probabilistic event. To ensure the quality of service, it is important to adopt time constraint to limit the efficiency of task completion, which is another factor considered to model in this paper. The workflow of the entire model can be summarized and displayed in the Fig.\ref{fig3.2}.

\subsection{DECN Scenario}
A DECN is composed of several deployment plans, which could be represented by $L=\{S,I,X,Y,P\}$, the detailed notations could be explained as:
\begin{itemize}
    \item $S$: the set of nodes in a deployment plan , where $s_i$ means the $i$-th node in the deployment plan ;
    \item $I$: the set of branches in a deployment plan , where $I_i$ means the $i$-th branch in the deployment plan ;
    \item $X$: the capacity of branches, where $X_i$ is the maximal capacity of bi, and $x_i$ is the actual capacity of $b_i$
    \item $Y$: the performance of nodes in a deployment plan , where $Y_i$ is the best performance that the $s_i$ can provide, and $y_i$ is the actual performance that the $s_i$ would provide.
    \item $P$: the changes in data size after computing nodes, where $p_i$ means the ratio of data output to input through the node
\end{itemize}
Once the above structure parameters are determined, a typical DECN can be established. A hierarchical distributed service or computing task would be reasonably divided and assigned to different servers and devices on one deployment plan, according to the geographical order and hierarchical structure of its sub-tasks. All nodes designed for the deployment plan will jointly complete the task. Each node that exists both input and output need extra time to finish receiving and sending work of data packets, named lead time,which would be attached to branches. For the convenience of calculation, no more details are given here. In fact, the delay of transmission is composed of many factors, but this study has simplified it here \cite{jiang2018low}.

Besides the lead time, the transmission for intermediate data of a model is more essential for the whole mission considering time constraints. For one deployment plan, the time required for transmission could be represented by the sum of the time on each branch. Equation \eqref{eq2} provides the answer of lead-time and transmission time Where $xi$ represents the available bandwidth on the branch  $i$ and $C$ is the initial input data. $L_i$ is the leadtime on the branch $i$. Rounding up is for calculation convenience.
\begin{equation}
    t=\sum_{i=1}^{n} L_{i}+\sum_{i=1}^{n}\left\lceil\frac{C}{x_{i}}\right\rceil\label{eq2}
\end{equation}
The calculation of $t$ so far is similar to traditional network transmission reliability research \cite{huang2020network}, because only the time it takes for data to be transmitted on the deployment plan has been considered. In DECN, once the data arrives at a node as input, the node would be activated to execute local task. At present, it is considered that the running time of the execution calculation of the copy running on each node is proportional to the size of the input data, and inversely proportional to the available computing resource. In addition, due to the unknowability of the tasks amount already running on the node, the performance of the node is considered in a dynamic situation, which is difficult to perceive available computing resource of the node when the computing task is coming. So changeable $yi$ like third term in equation \eqref{eq3} is used to represent the required computing resource.
\begin{equation}
    t=\sum_{i=1}^{n} L_{i}+\sum_{i=1}^{n}\left\lceil\frac{C}{x_{i}}\right\rceil+\sum_{i=1}^{n} \frac{C}{y_{i}} \label{eq3}
\end{equation}
Each $x_i$ and $y_i$ is independent, which means they can take any value among their range. However, the size of the output data after the computing node would not be exactly the same as the input data, such as  dimension reduction and convolution operations. The ratio of output to input of the calculation executed on the node would be saved as a parameter and used as data change factors. Combining the data size change of the node and an initial demand $C_0$, the time finally completing the whole task for the deployment plan can be expressed as:
\begin{equation}
    t=\sum_{i=1}^{n} L_{i}+\sum_{i=1}^{n}\left\lceil\frac{C_{0} \prod_{k=0}^{i-1} S_{k}}{x_{i}}\right\rceil+\sum_{i=0}^{n} \frac{C_{0} \prod_{k=0}^{i-1} S_{k}}{y_{i}}\label{eq4}
\end{equation}
Under the constraint of time T, if the entire computing task which takes $C_0$ as input can execute successfully, the inequality $t\leq T$ need be satisfied. The above formula is  OK under normal situation, but when the rate appears 0, attention need to be paid to it during the experiment. Because the node could have no computing task, the data must be transmitted.
\subsection{MSV and RSDP Algorithms}
For one specific deployment plan in DECN, the lead time is fixed first. Assuming that the solution $(x_i,y_i)=
    \{X_i(x_0,x_1,x_2, ...x_n), Y_i(y_0,y_1,y_2, ...y_n)\}$ satisfies $t\leq T$. Time variable $t$ is mostly determined by the transmission time and calculation time, affected by the available bandwidth of all deployment plans and  computing resource of all nodes. This solution $(x_i,y_i)$ is one of the feasible solutions for the reliability of the deployment plan, which could be answered by follow equation \eqref{eq5}:
\begin{equation}
    R_i=\operatorname{Pr}\{X, Y \mid t \leq T\}=\operatorname{Pr}\left(\prod_{i=1}^{n} x_i \prod_{i=1}^{n} y_i\right)\label{eq5}
\end{equation}
Due to the diversity of deployment plan bandwidth distribution and node computing resources distribution, there usually exists many solutions similar to the solutions$p_i$. Assuming that there are totally $z$ solutions in the set of  $\O_z$ where stores every solution for the inequality $t\leq T$, the global reliability of a single deployment plan can be defined as:
\begin{equation}
    R_{a}=\operatorname{Pr}\left\{\bigcup_{i=1}^{z} R_i\right\}\label{eq6}
\end{equation}
However, in this case the calculation method for reliability of $\O_z$ exists a lot of repeated calculations, bring a low efficiency. So, the Minimal Status Vectors (MSV) algorithm is used to eliminate all redundant calculation in this paper. Vector operations are depicted as the following principles:
\begin{itemize}
    \item Rule 1. $Y\leq X,(y_1,y_2,...,y_n)\leq(x_1,x_2,...,x_n):y_i\leq x_i$ for each $i=1, 2, ...,n$.
    \item Rule 2. $Y<X,(y_1,y_2,...,y_n)<(x_1,x_2,...,x_n):Y\leq $ and$y_i < x_i$for at least one $i$.
\end{itemize}

It would be removed for such kind of $X$ in $\O _z$, that the $X \geq $ any $Y$ (except for $X$) and $ \forall X,Y \in \O_z$. Then,  the remaining solutions are more representative, also named MSVs. Here is a proof that the reliability of distributed tasks on a certain deployment plan can be represented by the probability calculated by all MSVs on the plan.
\par

\begin{proof}
    $\forall {P_i, P_j} \in \O_z\ and\ P_i \geq P_j$, obeying above principles rule. \\
    \indent $P_i={X_i(x_0,x_1,x_2, ...x_n) Y_i(y_0,y_1,y_2, ...y_n)}$.\\
    \indent $P_j={X_j(x_0,x_1,x_2, ...x_n) Y_j(y_0,y_1,y_2, ...y_n)}$.\\
    \indent $Pr(P_i)=\prod_{k=0}^{n}p(x\geq x_k)p(y\geq y_k)$\\
    $because: P_i\geq P_j$, assuming that $X_i(x_0)\geq X_j(x_0)$\\
    \indent $\ and\ X_i(x_1,x_2, ...x_n)=X_j(x_1,x_2, ...x_n),\ Y_i=Y_j$\\
    $therefore: Pr(P_j) = Pr(P_i)+Pr\{X(x_0)=X_j(x_0)\}$ \\
    $therefore:$ \\
    $Pr(P_j) = Pr(P_i)+p\{x=X_i(x_0)\} \times \prod_{k=1}^{n}p(x\geq x_k)\ p(y\geq y_k)$ \\
    so $Pr\{P_i,P_j\}$\ exists duplicate calculations without properly process on the solution set. So MSVs method is used to  remove the solution set that will be recalculated.
\end{proof}
\par

If there exists $z$ MSVs in total in new answer set called $\O_{MSVs}$, the reliability of deployment plan change into \eqref{eq7}:
\begin{equation}
    R=\operatorname{Pr}\left\{\bigcup_{j=1}^{z} R j\right\}\label{eq7}
\end{equation}
The calculation of equation \eqref{eq7} is a joint probability problem of multiple probabilities. To avoid repeated calculations, the principle of inclusion and exclusion is a feasible idea for solving this kind of problem. The RSDP algorithm mentioned in paper \cite{zuo2007efficient} provides an efficient way using recursive thinking. In the case where all solutions $R_i=Pr\{\cdot\}$ can be found, the reliability of $\O_z$ which contains a total of $z$ solutions could be answered by $R=RSDP(\O_z)$.

\subsection{The REMR Algorithm}
The REMR means a reliability evaluation method with MSV and RSDP algorithm. Once the DECN is established, the input data and time constraint identified, the service reliability of a single and global deployment can be obtained by the Algorithm \ref{RA}. The line 25 gave the result of a single deployment plan, while the global reliability could be calculated by line 27.

The whole algorithm is mainly composed of several parts. The lines 4 to 10 is to find a solution set that meets the requirements, and lines 11 to 24 is to find all MSVs in this solution set. Then the RSDP method combined with the MSVs is used to solve the reliability of the corresponding deployment plan in line 25. Finally, the reliability of the entire scheme is calculated by joint probability formulas.

Assuming that there are a total of $m$ initial vector in the solution set, the complexity of the algorithm for finding MSVs is about $O(m^2)$. If there are a total of n MSVs for one deployment plan, the complexity of the RSDP method would be $O(n^3)$. The last step of whole algorithm, the joint probability only have $O(k)$ complexity if there exists $k$ plans in a DECN.

\begin{algorithm}[htb]
    \caption{Multi-Deployment Plans Reliability}
    \label{RA}
    \begin{algorithmic}[1] 
        \REQUIRE ~~\\ 
        Initial $DECN: S, I, X, Y, P$;\\
        Input $input\_data,\  T$;\\
        Solve: $\O_{MSVs}, RSDP(\O_{MSVs})$;
        \ENSURE ~~\\ 
        $R$: reliability of whole DECN;
        \STATE $t=\sum leadtime+trans time+computing time$;\
        \STATE Suppose there are K deployment plans;\
        \FOR{$k = 1:K$}
        \STATE Create a traversal table $\O_{init}$ with N vectors for {X,Y};\
        \STATE Create two empty collections $\O_z,\O_{MSVs}$
        \FOR{$n = 1:N$ in $\O_{init}$}
        \IF{$t\_n\leq T$}
        \STATE $\O_z = \O_z \cup {Vector:n}$\
        \ENDIF
        \ENDFOR
        \STATE Suppose there are $Z$ solutions in $\O_z$, $L$ MSVs in $\O_{MSVs}$;
        \FOR{$z = 1:Z$ in $\O_{z}$}
        \FOR{$l = 1:L$ in $\O_{MSVs}$}
        \IF{$MSV:l \geq Solution: z  $}
        \STATE Remove $MSV: l$ from $\O_{MSVs}$;\
        \ENDIF
        \ENDFOR
        \FOR{$l = 1:L$ in $\O_{MSVs}$}
        \IF{$Solution: z\geq MSV:l$}
        \STATE Break and next $z$;\
        \ENDIF
        \ENDFOR
        \STATE $\O_{MSVs} = \O_{MSVs} \cup solution: z$
        \ENDFOR
        \STATE $R_k=RSDP(\O_{MSVs})$;\
        \ENDFOR
        \STATE $R = \bigcup_{k=1}^{K} R_k$;\
    \end{algorithmic}
\end{algorithm}

\section{Experiments in DECN}
\begin{figure}[htbp]
    \centerline{\includegraphics[width=0.5\textwidth]{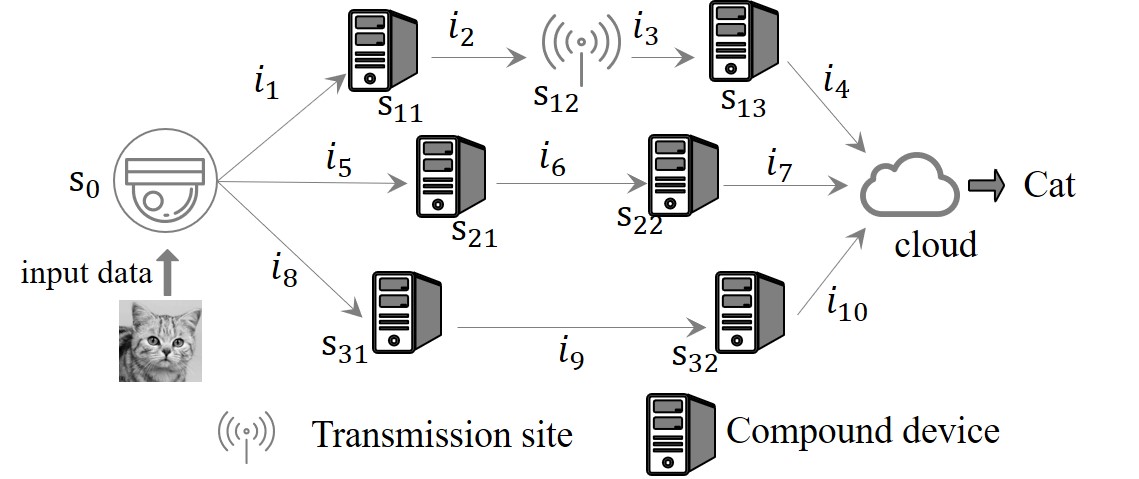}}
    \caption{An illustrative DECN example}
    \label{fig4.1}
\end{figure}

A DECN designed like Fig.\ref{fig4.1} is used to explain the entire solution procedure where the deployment plan of the service have been shown. In this example,  $S_0$ node means  a edge device where the task data from and the $cloud$ is the terminal for the deployment plan. The compound nodes will play a role in transmission task and computing task simultaneously, while the transport nodes will only account for transmission task.

\begin{figure}[htbp]
    \centerline{\includegraphics[width=0.5\textwidth]{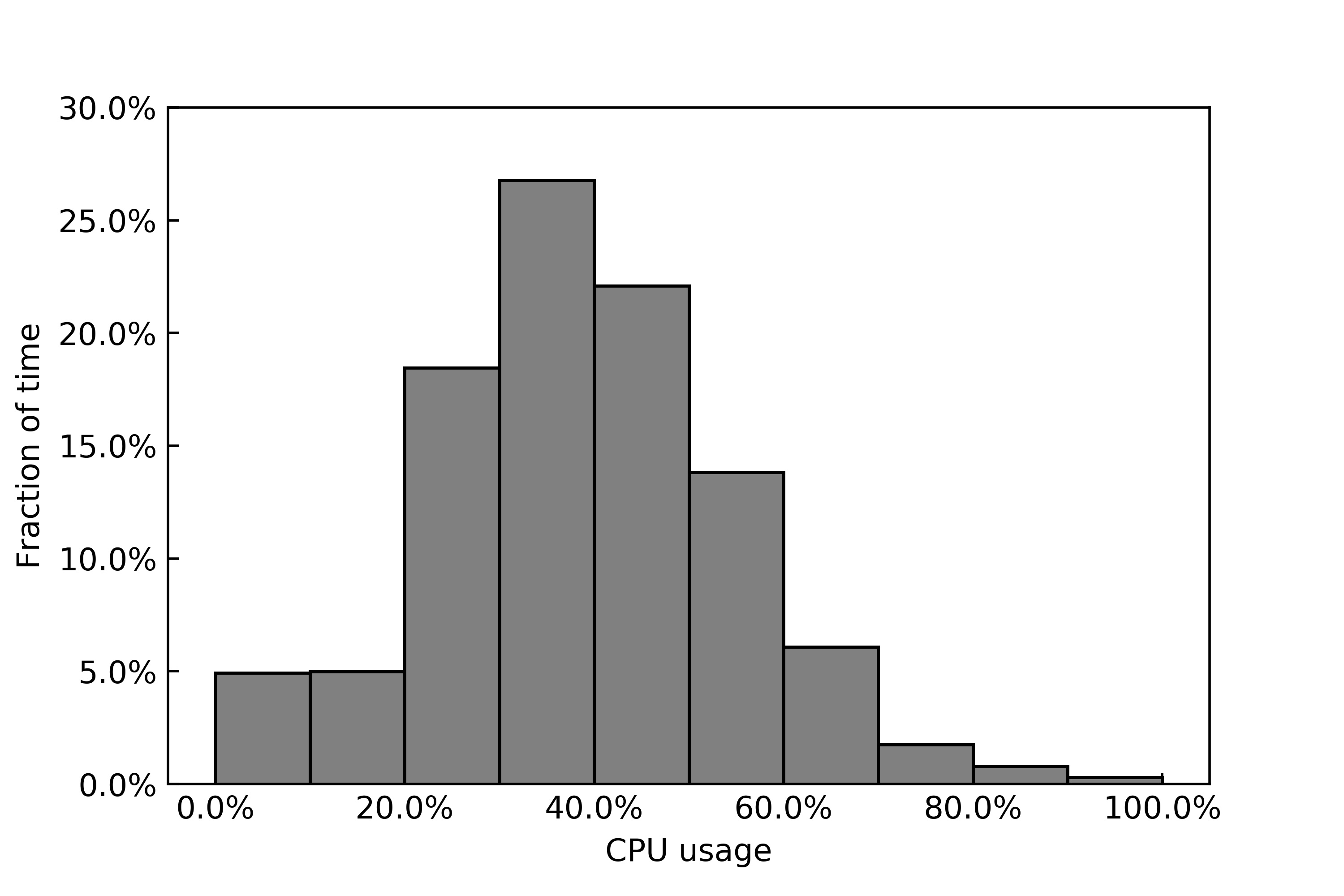}}
    \caption{CPU usage distribution with time in Alibaba servers cluster.}
    \label{fig4.2}
\end{figure}

The TABLE \ref{tab1} depicts the probability distribution of available bandwidth on each branch and the TABLE  \ref{tab2} shows the probability distribution of available compute resources on each server and each device, while each capacity of bandwidth and compute resources is set to be a multiple of integer for simplicity. The bandwidth distribution based on historical data is derived from paper \cite{huang2016routing}, used to describe the probability of each branch's available bandwidth. Then none of papers in network reliability study the computing resources. The distribution of computing resources TABLE \ref{tab2} in this paper comes from the analysis of the Alibaba servers dataset. Alibaba provides a huge dataset, which includes the log records of CPU usage about server clusters \cite{9068614}. This study used the data to simulate the CPU usage frequency of the edge server and converted it into available CPU resources, as shown in the Fig. \ref{fig4.2}. The $output/input$ in the TABLE \ref{tab2} represented the ratio of the output data calculated by the server to the size of the input data received by the server, which is used to calculate the change degree in the deployment plan.

\begin{table}[htbp]
    \caption{Bandwidth Capacity Probability}
    \begin{center}
        \begin{tabular}{cccccccc}
            \hline
            \multirow{2}{*}{branch} & \multirow{2}{*}{lead time} & \multicolumn{6}{l}{capacity}                                    \\ \cline{3-8}
                                    &                            & 0                            & 1    & 2    & 3    & 4    & 5    \\ \hline
            $i_1$                   & 2                          & 0.04                         & 0.01 & 0    & 0.02 & 0    & 0.93 \\
            $i_2$                   & 1                          & 0.03                         & 0.01 & 0.02 & 0.94 & 0    & 0    \\
            $i_3$                   & 1                          & 0.02                         & 0.01 & 0    & 0.04 & 0.93 & 0    \\
            $i_4$                   & 1                          & 0.03                         & 0.01 & 0    & 0.06 & 0    & 0.9  \\
            $i_5$                   & 2                          & 0.02                         & 0.01 & 0.01 & 0    & 0.96 & 0    \\
            $i_6$                   & 1                          & 0.04                         & 0.01 & 0.95 & 0    & 0    & 0    \\
            $i_7$                   & 1                          & 0.03                         & 0.01 & 0    & 0.02 & 0.94 & 0    \\
            $i_8$                   & 1                          & 0.03                         & 0.01 & 0.01 & 0.95 & 0    & 0    \\
            $i_9$                   & 2                          & 0.03                         & 0    & 0.01 & 0.03 & 0.93 & 0    \\
            $i_{10}$                & 1                          & 0.02                         & 0    & 0.01 & 0.02 & 0.95 & 0    \\ \hline
        \end{tabular}
        \label{tab1}
    \end{center}
\end{table}

\begin{table}[htbp]
    \caption{Available Nodes Capacity Probability}
    \begin{center}
        \begin{tabular}{ccccccccc}
            \hline
            \multirow{2}{*}{nodes} & \multirow{2}{*}{$\frac{output}{input}$} & \multicolumn{7}{l}{capacity}                                           \\ \cline{3-9}
                                   &                                         & 0                            & 1    & 2    & 3    & 4    & 5    & 6    \\ \hline
            $s_0$                  & 0.8                                     & 0.0                          & 0.01 & 0.09 & 0.26 & 0.37 & 0.20 & 0.07 \\
            $s_{11}$               & 1.2                                     & 0.0                          & 0.01 & 0.09 & 0.26 & 0.37 & 0.20 & 0.07 \\
            $s_{12}$               & 1.0                                     & 0.0                          & 0.01 & 0.09 & 0.26 & 0.37 & 0.20 & 0.07 \\
            $s_{13}$               & nan                                     & 0.0                          & 0.01 & 0.09 & 0.26 & 0.37 & 0.20 & 0.07 \\
            $s_{21}$               & 1.2                                     & 0.0                          & 0.01 & 0.09 & 0.26 & 0.37 & 0.20 & 0.07 \\
            $s_{22}$               & nan                                     & 0.0                          & 0.01 & 0.09 & 0.26 & 0.37 & 0.20 & 0.07 \\
            $s_{31}$               & 1.2                                     & 0.0                          & 0.01 & 0.09 & 0.26 & 0.37 & 0.20 & 0.07 \\
            $s_{32}$               & nan                                     & 0.0                          & 0.01 & 0.09 & 0.26 & 0.37 & 0.20 & 0.07 \\ \hline
        \end{tabular}
    \end{center}
    \label{tab2}
\end{table}

\subsection{An Illustrative Example}
Assuming that $C_0=15$ units of data would be the input of the three deployment plans with time constraint $T = 25s$, the reliability of the task could be present by $R=\{R_a, R_b, R_c\}$, where $a=\{s_0,s_{12},s_{13},s_{11},cloud\}, b=\{s_0,s_{21},s_{22},cloud\}, c=\{s_0,s_{31},s_{32},cloud\}$. To explain this indicator $output/input$ with more detail, TABLE \ref{tab3} is draw where the DP represented different deployment lines. They have one initial edge device in common. Therefore, the initial input $C_0$ is all 15 units. Taking DP1 as an example, The indicator of the device is 0.8, so the amount of data send to the first node $C_1$of DP1 becomes 12. Here the indicator is 1.2, so the amount of data $C_2$ becomes 14.4. Since the $s_{12}$ in DP1 is a transmission node without computing, the output $C_3$ is totally the same as input, also 14.4, the node $S_{13}$ would finish the final computing task and send a result to the  cloud. Usually the output here would be very small, so the indicator is set to be 0.01, to present a tiny computing task.
This section would firstly take $R_a$ as the example to describe specific calculation process. According to the equation \eqref{eq7}, all MSVs need to be found. Combining the equation \eqref{eq4}, the given initial $input\_data$ and time threshold $T$, implicit enumeration method is utilized in the above-mentioned DECN to gain all solutions for the inequality $t=4+\sum_{i=1}^{4}\left\lceil\frac{C_0 \prod_{k=0}^{i-1} S_{k}}{x_{i}}\right\rceil+\sum_{i=0}^{4} \frac{C_0 \prod_{k=0}^{i} S_{k}}{y_{i}}\leq 25s$ without thinking of repetition. The changes in the amount of input data after node processing are shown in the table \ref{tab3}. $C_0$ represents the initial input, and the amount of data would always change after going through different nodes of the deployment plan, similarly $C_i$ represents the amount of data output after the $i$-th node, and the rest is the same. Implicit enumeration refers to the removal of attempts to impossible answers on the basis of enumeration. First 2460 solutions were found in solving the inequality. After filtering the solution set, 14 MSVs remained, while part of them were listed in the table \ref{tab4}.

\begin{table}[htbp]
    \caption{Changes of Data Size in Deployment Plans}
    \begin{center}
        \setlength{\tabcolsep}{5mm}{
            \begin{tabular}{cccllll}
                \hline
                    & $C_0$ & $C_1$ & $C_2$ & $C_3$ & $C_4$ \\ \hline
                DP1 & 15    & 12    & 14.4  & 14.4  & 0.01  \\
                DP2 & 15    & 12    & 14.4  & 0.01  &       \\
                DP3 & 15    & 12    & 14.4  & 0.01  &       \\ \hline
            \end{tabular}}
        \label{tab3}
    \end{center}
\end{table}

\begin{table}[htbp]
    \caption{The MSVs for A Deployment Plan}
    \begin{center}
        \setlength{\tabcolsep}{6mm}{
            \begin{tabular}{ccclllll}
                \hline
                \multirow{2}{*}{No}    & \multirow{2}{*}{$X(x1,x2,x3,x4)$} & \multicolumn{6}{c}{\multirow{2}{*}{$Y(y1,y2,y3)$}} \\
                                       &                                   & \multicolumn{6}{c}{}                               \\ \hline
                1                      & (2, 3, 4, 1)                      & \multicolumn{6}{c}{(5, 1, 6)}                      \\
                2                      & (2, 3, 4, 1)                      & \multicolumn{6}{c}{(6, 1, 5)}                      \\
                3                      & (3, 3, 3, 1)                      & \multicolumn{6}{c}{(4, 1, 5)}                      \\
                4                      & (3, 3, 3, 1)                      & \multicolumn{6}{c}{(5, 1, 4)}                      \\
                5                      & (3, 3, 4, 1)                      & \multicolumn{6}{c}{(4, 1, 4)}                      \\
                .                      & .                                 & \multicolumn{6}{c}{.}                              \\
                .                      & .                                 & \multicolumn{6}{c}{.}                              \\
                .                      & .                                 & \multicolumn{6}{c}{.}                              \\
                \multicolumn{1}{l}{14} & (4, 3, 4, 1)                      & \multicolumn{6}{c}{(4, 1, 3)}                      \\ \hline
            \end{tabular}}
        \label{tab4}
    \end{center}
\end{table}

Then the reliability of deployment plan $a$ can be obtained by the improved RSDP algorithm. The reliability $R_a=0.82344$. The reliability of $R_b = 0.87679$ and $R_c = 0.90955$ can also be get by same way above,  while they respectively own 7 and 17 MSVs. Finally the global reliability $R = R_a\bigcup R_b\bigcup R_c$. Considering that the three reliability are independent, so
\begin{equation}
    R = R_a + R_b*(1-R_a)+R_c*(1-R_a)*(1-R_b)\label{eq8}
\end{equation}
The obtained result is $0.99653$, which presents the probability of the DECN finishing the task perfectly.
More similar experiments based on the same DECN were carried out and the results were record at the TABLE \ref{tab5} below.

\begin{table}[htbp]
    \caption{ Numerical Experiments}
    \begin{center}
        \setlength{\tabcolsep}{6mm}{
            \begin{tabular}{ccll}
                \hline
                                         & C=14    & C=15    & C=16    \\ \hline
                T=20                     & 0.94930 & 0.87567 & 0.75312 \\
                T=21                     & 0.97163 & 0.94656 & 0.82634 \\
                T=22                     & 0.98670 & 0.97449 & 0.91367 \\
                T=24                     & 0.99606 & 0.99438 & 0.98362 \\
                \multicolumn{1}{l}{T=25} & 0.99729 & 0.99653 & 0.99207 \\ \hline
            \end{tabular}}
        \label{tab5}
    \end{center}
\end{table}

\subsection{Experiments on Google Cluster Trace Dataset}
\begin{figure}[htbp]
    \centerline{\includegraphics[width=0.5\textwidth]{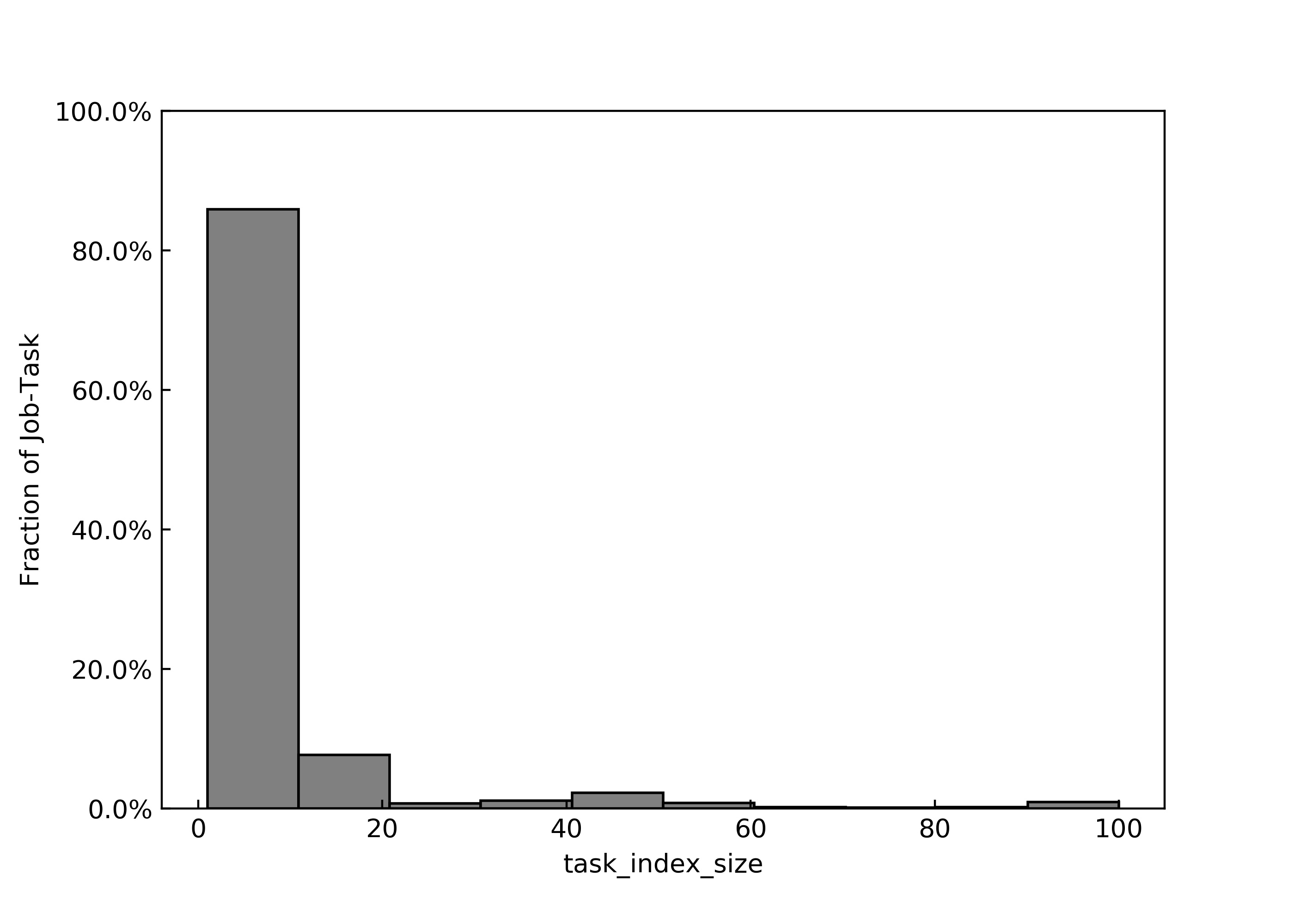}}
    \caption{the Distribution of subtasks number in part Google Cluster Trace Dataset}
    \label{fig4.3}
\end{figure}

\begin{figure}[htbp]
    \centerline{\includegraphics[width=0.5\textwidth]{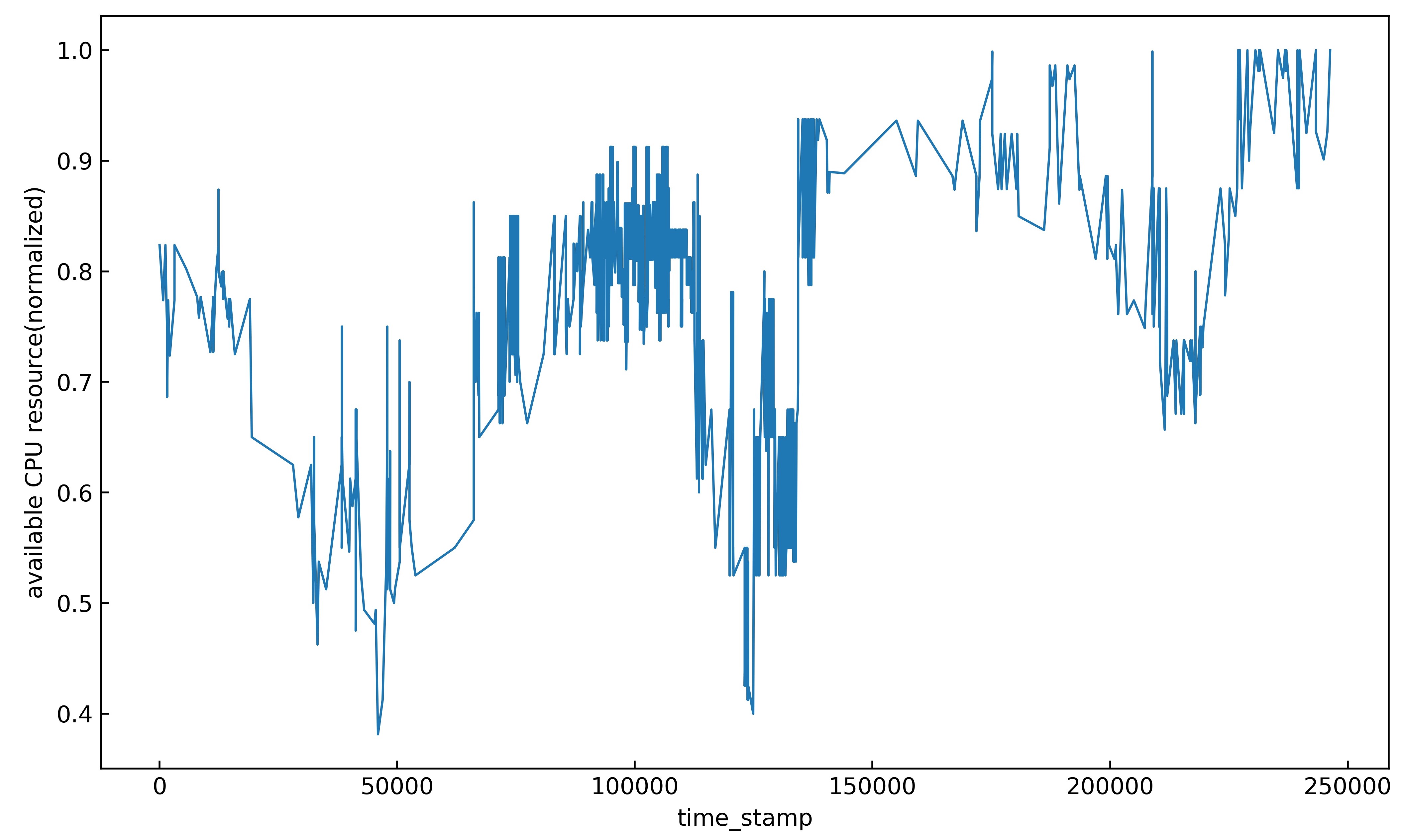}}
    \caption{the Curve of available CPU resource with time in Google Cluster Trace Data-set.}
    \label{fig4.4}
\end{figure}

\begin{figure}[htbp]
    \centerline{\includegraphics[width=0.5\textwidth]{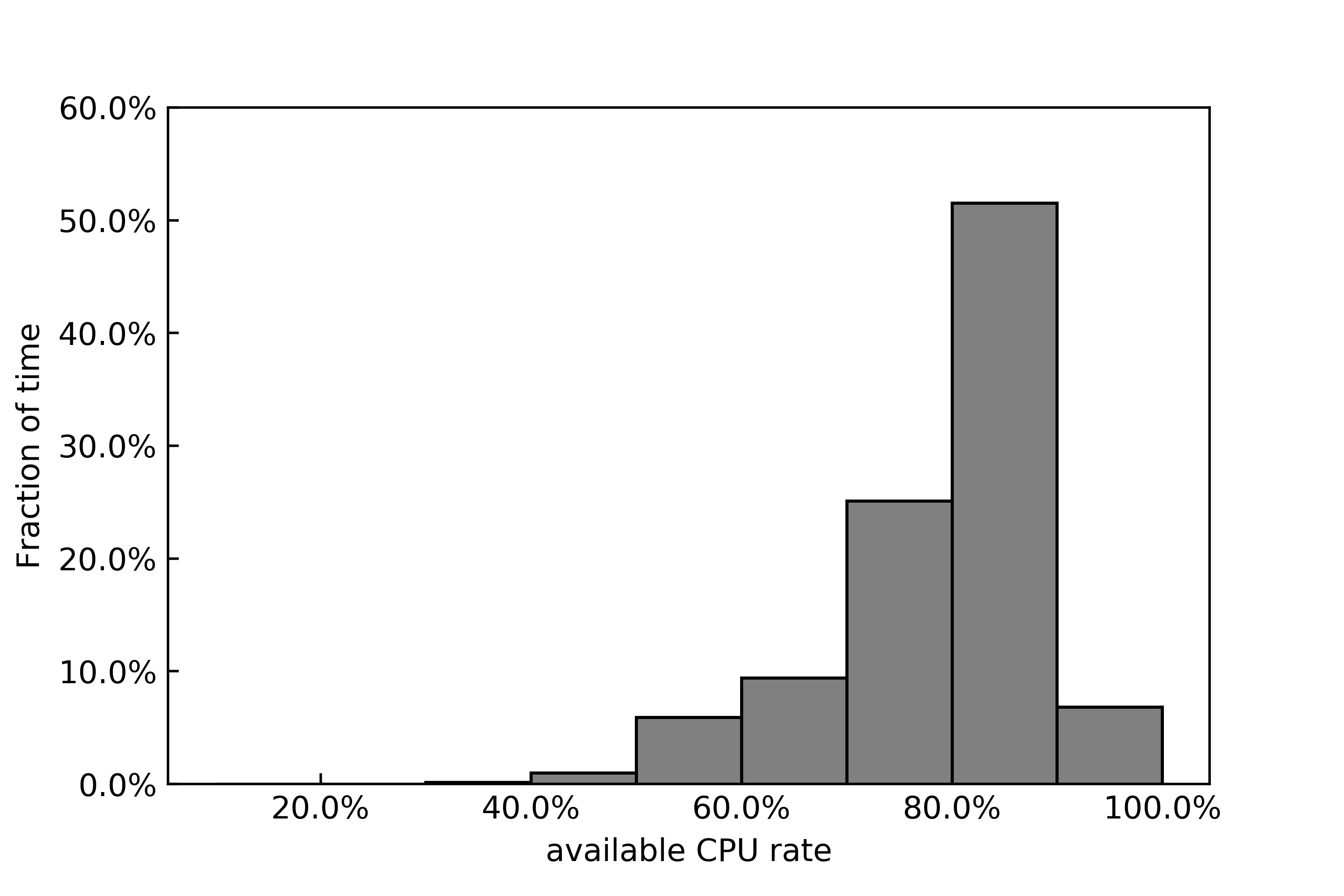}}
    \caption{available CPU resource distribution in part of Google Cluster Trace Data-set.}
    \label{fig4.5}
\end{figure}

The google cluster data trace represents 29 day's worth of Borg cell information from May 2011, on a cluster of about 12.5k machines. According to the analysis of the data-set in this study, it was found that most of the services were decomposed into several sub-services. The distribution of the  sub-services number can be statistically obtained as shown in Fig.\ref{fig4.3} by counting the frequency of event type 0 in each job. Obviously most of the services are decomposed into no more than 10 sub-services. This experiment selected a service with 3 sub tasks in the data-set for evaluation. These three sub-services run on different machines based on the time sequence, so this study monitored the available resources on these three machines based on three days of data. Fig.\ref{fig4.4} showed the situation of one machine of the three, and it was translated into a probability distribution shown in Fig.\ref{fig4.5}. Since the cluster data-set have no bandwidth distribution data, the bandwidth data-set used in last experiment was adopted again. The reliability is calculated to be 0.98657, while the input data size is 14 and time constraint is 20. Although possibly affected by the bandwidth, the success rate of its task completion is still very high which could be explained by the high spare rate of machines.

\section{Conclusions}
Changeable factors including computing resources, transmission bandwidth and intermediate data volume make the previous reliability evaluation methods less accurate. This paper provides a new reliability evaluation method by establishing a DECN scenario under time constraint. Aiming to present the dynamic scenario of edge computing with distributed services, the built model have edge devices and edge servers, owning different probability distribution to describe their respective capacity. Through the MSVs and RSDP algorithm, the reliability based on solution set could be obtained. Then an example was introduced to explain how the model and method work. At last, NS3 was used to experiment as the simulation, which is in line with the proposed method almost. With the method, it would be convenient to tell how reliable a distributed service is and how to make the service more reliable.

Some directions worthy of further research in the future are listed as follows. The first is to optimize the time calculation way when the nodes executing tasks, while the current evaluation method is not detailed enough. Then the RSDP Algorithm could be improved on time as showed in paper \cite{bai2015ordering}. More importantly,  hoping to provide a framework, so that the proposed reliability evaluation method could be used to segment a distributed service or a DNN model rather than manual assignments.

\bibliography{references} 

\begin{thebibliography}{10}

\bibitem{top10strategic}
G.~Top, ``Strategic technology trends for 2020,'' {\em Smarter with
  Gartner.--URL: https://www. gartner.
  com/smarterwithgartner/gartner-top-10-strategic-technologytrends-for-2020},
  10.

\bibitem{shirer2019growth}
M.~Shirer and C.~MacGillivray, ``The growth in connected iot devices is
  expected to generate 79.4 zb of data in 2025, according to a new idc
  forecast,'' IDC, 2019.

\bibitem{nord2019internet}
J.~H. Nord, A.~Koohang, and J.~Paliszkiewicz, ``The internet of things: Review
  and theoretical framework,'' {\em Expert Systems with Applications},
  vol.~133, pp.~97--108, 2019.

\bibitem{8736011}
Z.~Zhou, X.~Chen, E.~Li, L.~Zeng, K.~Luo, and J.~Zhang, ``Edge intelligence:
  Paving the last mile of artificial intelligence with edge computing,'' {\em
  Proceedings of the IEEE}, vol.~107, no.~8, pp.~1738--1762, 2019.

\bibitem{robinson2014using}
G.~Robinson, I.~Vamvadelis, A.~Narin, {\em et~al.}, ``Using amazon web services
  for disaster recovery,'' {\em Amazon web services}, vol.~22, 2014.

\bibitem{shi2016edge}
W.~Shi, J.~Cao, Q.~Zhang, Y.~Li, and L.~Xu, ``Edge computing: Vision and
  challenges,'' {\em IEEE internet of things journal}, vol.~3, no.~5,
  pp.~637--646, 2016.

\bibitem{8030322}
N.~Abbas, Y.~Zhang, A.~Taherkordi, and T.~Skeie, ``Mobile edge computing: A
  survey,'' {\em IEEE Internet of Things Journal}, vol.~5, no.~1, pp.~450--465,
  2018.

\bibitem{yi2015survey}
S.~Yi, C.~Li, and Q.~Li, ``A survey of fog computing: concepts, applications
  and issues,'' in {\em Proceedings of the 2015 workshop on mobile big data},
  pp.~37--42, 2015.

\bibitem{king2016distributed}
J.~King and A.~I. Awad, ``A distributed security mechanism for
  resource-constrained iot devices,'' {\em Informatica}, vol.~40, no.~1, 2016.

\bibitem{mao2017survey}
Y.~Mao, C.~You, J.~Zhang, K.~Huang, and K.~B. Letaief, ``A survey on mobile
  edge computing: The communication perspective,'' {\em IEEE Communications
  Surveys \& Tutorials}, vol.~19, no.~4, pp.~2322--2358, 2017.

\bibitem{simonyan2014very}
K.~Simonyan and A.~Zisserman, ``Very deep convolutional networks for
  large-scale image recognition,'' {\em arXiv preprint arXiv:1409.1556}, 2014.

\bibitem{fu2021adaptive}
K.~Fu, W.~Zhang, Q.~Chen, D.~Zeng, and M.~Guo, ``Adaptive resource efficient
  microservice deployment in cloud-edge continuum,'' {\em IEEE Transactions on
  Parallel and Distributed Systems}, 2021.

\bibitem{satyanarayanan2017emergence}
M.~Satyanarayanan, ``The emergence of edge computing,'' {\em Computer},
  vol.~50, no.~1, pp.~30--39, 2017.

\bibitem{9272869}
S.~Long, W.~Long, Z.~Li, K.~Li, Y.~Xia, and Z.~Tang, ``A game-based approach
  for cost-aware task assignment with qos constraint in collaborative edge and
  cloud environments,'' {\em IEEE Transactions on Parallel and Distributed
  Systems}, vol.~32, no.~7, pp.~1629--1640, 2021.

\bibitem{milocco2020evaluating}
R.~Milocco, P.~Minet, E.~Renault, and S.~Boumerdassi, ``Evaluating the upper
  bound of energy cost saving by proactive data center management,'' {\em IEEE
  Transactions on Network and Service Management}, vol.~17, no.~3,
  pp.~1527--1541, 2020.

\bibitem{huang2020network}
C.-F. Huang, D.-H. Huang, and Y.-K. Lin, ``Network reliability evaluation for a
  distributed network with edge computing,'' {\em Computers \& Industrial
  Engineering}, vol.~147, p.~106492, 2020.

\bibitem{8576668}
B.~Wang, Y.~Lei, N.~Li, and N.~Li, ``A hybrid prognostics approach for
  estimating remaining useful life of rolling element bearings,'' {\em IEEE
  Transactions on Reliability}, vol.~69, no.~1, pp.~401--412, 2020.

\bibitem{8776630}
D.-H. Huang, C.-F. Huang, and Y.-K. Lin, ``A binding algorithm of lower
  boundary points generation for network reliability evaluation,'' {\em IEEE
  Transactions on Reliability}, vol.~69, no.~3, pp.~1087--1096, 2020.

\bibitem{8932578}
S.~Chakraborty, N.~K. Goyal, S.~Mahapatra, and S.~Soh, ``Minimal path-based
  reliability model for wireless sensor networks with multistate nodes,'' {\em
  IEEE Transactions on Reliability}, vol.~69, no.~1, pp.~382--400, 2020.

\bibitem{7964780}
O.~Kabadurmus and A.~E. Smith, ``Evaluating reliability/survivability of
  capacitated wireless networks,'' {\em IEEE Transactions on Reliability},
  vol.~67, no.~1, pp.~26--40, 2018.

\bibitem{9468876}
C.-F. Huang, D.-H. Huang, and Y.-K. Lin, ``Reliability evaluation of a
  cloud–fog computing network considering transmission mechanisms,'' {\em
  IEEE Transactions on Reliability}, pp.~1--13, 2021.

\bibitem{loh2011addressing}
R.~C. Loh, S.~Soh, and M.~Lazarescu, ``Addressing the most reliable
  edge-disjoint paths with a delay constraint,'' {\em IEEE Transactions on
  Reliability}, vol.~60, no.~1, pp.~88--93, 2011.

\bibitem{huang2021novel}
C.-H. Huang, D.-H. Huang, and Y.-K. Lin, ``A novel approach to predict network
  reliability for multistate networks by a deep neural network,'' {\em Quality
  Technology \& Quantitative Management}, pp.~1--17, 2021.

\bibitem{ramirez2005monte}
J.~E. Ramirez-Marquez and D.~W. Coit, ``A monte-carlo simulation approach for
  approximating multi-state two-terminal reliability,'' {\em Reliability
  Engineering \& System Safety}, vol.~87, no.~2, pp.~253--264, 2005.

\bibitem{dong2019reliability}
L.~Dong, W.~Wu, Q.~Guo, M.~N. Satpute, T.~Znati, and D.~Z. Du,
  ``Reliability-aware offloading and allocation in multilevel edge computing
  system,'' {\em IEEE Transactions on Reliability}, 2019.

\bibitem{li2019edge}
E.~Li, L.~Zeng, Z.~Zhou, and X.~Chen, ``Edge ai: On-demand accelerating deep
  neural network inference via edge computing,'' {\em IEEE Transactions on
  Wireless Communications}, vol.~19, no.~1, pp.~447--457, 2019.

\bibitem{kang2017neurosurgeon}
Y.~Kang, J.~Hauswald, C.~Gao, A.~Rovinski, T.~Mudge, J.~Mars, and L.~Tang,
  ``Neurosurgeon: Collaborative intelligence between the cloud and mobile
  edge,'' {\em ACM SIGARCH Computer Architecture News}, vol.~45, no.~1,
  pp.~615--629, 2017.

\bibitem{jiang2018low}
X.~Jiang, H.~Shokri-Ghadikolaei, G.~Fodor, E.~Modiano, Z.~Pang, M.~Zorzi, and
  C.~Fischione, ``Low-latency networking: Where latency lurks and how to tame
  it,'' {\em Proceedings of the IEEE}, vol.~107, no.~2, pp.~280--306, 2018.

\bibitem{zuo2007efficient}
M.~J. Zuo, Z.~Tian, and H.-Z. Huang, ``An efficient method for reliability
  evaluation of multistate networks given all minimal path vectors,'' {\em IIE
  transactions}, vol.~39, no.~8, pp.~811--817, 2007.

\bibitem{huang2016routing}
C.-F. Huang, Y.-K. Lin, and L.~C.-L. Yeng, ``Routing scheme of a multi-state
  computer network employing a retransmission mechanism within a time
  threshold,'' {\em Information sciences}, vol.~340, pp.~321--336, 2016.

\bibitem{9068614}
J.~Guo, Z.~Chang, S.~Wang, H.~Ding, Y.~Feng, L.~Mao, and Y.~Bao, ``Who limits
  the resource efficiency of my datacenter: An analysis of alibaba datacenter
  traces,'' in {\em 2019 IEEE/ACM 27th International Symposium on Quality of
  Service (IWQoS)}, pp.~1--10, 2019.

\bibitem{bai2015ordering}
G.~Bai, M.~J. Zuo, and Z.~Tian, ``Ordering heuristics for reliability
  evaluation of multistate networks,'' {\em IEEE Transactions on Reliability},
  vol.~64, no.~3, pp.~1015--1023, 2015.

\end{thebibliography}
\bibliographystyle{ieeetr} 
\end{document}